\documentclass[conference]{IEEEtran}
\ifCLASSINFOpdf
\else
\fi
\usepackage{amsfonts}

\newcommand\mbbm[1]{\mathchoice{\mbox{\boldmath{$\displaystyle #1$}}}%
		{\mbox{\boldmath{$\textstyle #1$}}}%
		{\mbox{\boldmath{$\scriptstyle #1$}}}%
		{\mbox{\boldmath{$\scriptscriptstyle #1$}}}}%

\newcommand{\bF}{\mbbm{F}}

\newcommand{\blambda}{\mbbm{\lambda}}

\newcommand{\bx}{\mbbm{x}}

\newcommand{\by}{\mbbm{y}}

\newcommand{\mR}{\mathbb{R}}

\usepackage[dvips,dvipdfm]{graphicx}

\begin{document}
%
\title{A signal recovery algorithm for sparse matrix based compressed sensing}

\author{\IEEEauthorblockN{Yoshiyuki Kabashima}
\IEEEauthorblockA{
Dept. of Comput. Intel. and Syst. Sci. \\
Tokyo Institute of Technology\\
Yokohama, 226-8502, Japan\\
Email: kaba@dis.titech.ac.jp}
\and
\IEEEauthorblockN{Tadashi Wadayama}
\IEEEauthorblockA{
Dept. of Computer Science \\
Nagoya Institute of Technology\\
Nagoya, 466-8555, Japan \\
Email: wadayama@nitech.ac.jp}
}


%


\maketitle

\begin{abstract}
We have developed an approximate signal recovery algorithm with low computational cost for compressed sensing on the basis of randomly constructed sparse measurement matrices. The law of large numbers and the central limit theorem suggest that the developed algorithm saturates the Donoho-Tanner weak threshold for the perfect recovery when the matrix becomes as dense as the signal size $N$ and the number of measurements $M$ tends to infinity keep $\alpha=M/N \sim O(1)$, which is supported by extensive numerical experiments. Even when the numbers of non-zero entries per column/row in the measurement matrices are limited to $O(1)$, numerical experiments indicate that the algorithm can still typically recover the original signal perfectly with an $O(N)$ computational cost per update as well if the density $\rho$ of non-zero entries of the signal is lower than a certain critical value $\rho_{\rm th}(\alpha)$ as $N,M \to \infty$. 
\end{abstract}


%
\IEEEpeerreviewmaketitle

\section{Introduction}
Compressed (compressive) sensing (CS) is a framework that enables the recovery of a sparse signal from few of its measurements by exploiting the sparsity as the prior knowledge of the original signal. Famous applications of CS include computational photography and seismic data processing, in which the original signal and its measurements are linearly related, in most cases, by the Fourier and/or wavelet transformations. In general, these relationships are expressed by {\em dense} matrices. Accordingly, much effort has been made in analyzing the performance \cite{Donoho2006,CandesTao2005,Rangan2009,KWT2009} and 
developing 
practical algorithms 
for the signal recovery \cite{Blumensath2008,MalekiDonoho2009,DonohoMalekiMontanari2009} for the density matrix based CS. 

Data compression, data stream processing, and group testing are also included in examples of CS but may be classified into 
another category. Unlike the above applications, these examples naturally allow situations where each measurement is related to only a few entries of the original signal. 
In such cases, it is more suitable to model the signal-measurements relationship by not density but {\em sparse} matrices \cite{GilbertIndyk2010}. 

The $l_1$-norm minimization under the constraint of given measurements is widely adopted for the signal recovery of CS
and the interior point method is a standard solution for this task. In general, the computational cost per update of this scheme 
enlarges as cubic of the problem size, which is regarded as feasible in computational complexity theory. However, {\em in practice}, the cost may exceed the allowable limit depending on the problem size and/or situations. In such cases, the sparsity of the matrices can be advantageous for reducing the computational cost because necessary computations for relating the signal and measurements increase just {\em linearly} with respect to the signal size if the numbers of non-zero entries per column/row in the matrix are $O(1)$ \cite{GilbertIndyk2010,Akcakaya2009}. 

From this perspective, we have developed an {\em approximate} algorithm for the signal recovery of the sparse matrix based CS. In a similar setting, an earlier study \cite{CSBP2006} developed an algorithm on the basis of belief propagation (BP) \cite{Pearl1987,MacKay1998} in conjunction with using a mixture of two finite variance Gaussians as a sparse prior. The computational cost of the algorithm increases only linearly with respect to the signal size. However, the following two issues may be problematic: 
\begin{itemize}
\item Functions of continuous variables must be updated to deal with continuous signals. This means that the prefactor of the computational cost is rather large even if it is proportional to the signal size. 
\item The finiteness of the variances of the two Gaussians in the prior does not allow the perfect recovery of the original signal even if no noise is added to the measurements. 
\end{itemize}
We will show that these possible drawbacks can be resolved by using BP to the constrained $l_1$-norm minimization in conjunction with further quadratic approximation. We will also show that our algorithm asymptotically saturates the Donoho-Tanner weak threshold for the perfect recovery \cite{DonohoTanner2009} when the matrix becomes dense. This indicates that, in the dense limit, the algorithm developed here is as capable as the {\em approximate message passing} (AMP) proposed by Donoho et al. 
\cite{DonohoMalekiMontanari2009} 
although control of a soft-thresholding parameter is not necessary unlike AMP. 

The remainder of this article is organized as follows. The next section introduces the problem on which we will focus. Section III, which is the main part of this article, describes the suitable signal recovery algorithm for the sparse matrix based CS that we developed. Section IV examines the properties of the algorithm by extensive numerical experiments. The final section summaries the paper. 

\section{The problem setting}
In the general scenario, we suppose that each entry $x_i^0 \in \mR$ of an $N$-dimensional original signal $\bx^0=(x_i^0) \in \mR^N$ is independently generated from an identical distribution $P(x)=(1-\rho)\delta(x)+\rho \widetilde{P}(x)$. Here, $\widetilde{P}(x)$ is a certain distribution that has a finite variance but not a finite mass at the origin, and $\rho$ represents the density of non-zero entries of the signal. Multiplying $M(<N) \times N$ matrix $\bF=(F_{\mu i}) \in \mR^{M \times N}$ yields the linear measurements $\by=\bF \bx^0$ of dimensionality $M$. To simply characterize the sparsity of the measurement matrix, we focus on the case of {\em regular} ensembles in which positions of non-zero entries in $\bF$ are determined randomly under the constraint that the numbers of non-zero entries per column and row are fixed to finite values $j$ and $k$, respectively while $N, M \to \infty $ with keeping $\alpha=M/N \sim O(1)$. However, extension to {\em irregular} ensembles for which values of $j$ and $k$ are distributed in a matrix is straightforward. After the positions of the non-zero entries are fixed, each value is provided by independently sampling a random number from an identical distribution of a finite variance. 

To recover the original signal $\bx^0$ given $\by$ and $\bF$, we follow the constrained $l_1$-norm minimization scheme
\begin{eqnarray}
\mathop{\rm minimize} _{\bx}
\left \{
\sum_{i=1}^N |x_i| 
\right \}
\ \ 
{\rm subject \  to} 
\ \ 
\bF\bx=\by. 
\label{L1min}
\end{eqnarray}
This can be expressed as a problem of linear programming. In general, necessary computational cost for solving this scales as $O(N^3)$ per update by using the interior point method as a certain matrix inversion is required for each iteration \cite{interior}. The purpose of our study is to reduce this cost by developing an efficient approximate algorithm utilizing the framework of BP. 

\section{Algorithm development}
\subsection{Belief propagation}
As a basis of our study, we first convert eq. (\ref{L1min}) to an unconstrained optimization problem 
\begin{eqnarray}
\mathop{\rm max} _{\blambda}
\mathop{\rm min} _{\bx}
\left \{
\blambda^{\rm T} (\bF\bx-\by)+\sum_{i=1}^N |x_i| 
\right \}
\label{unconstrained}
\end{eqnarray}
by introducing the Lagrange multiplier $\blambda=(\lambda_\mu) \in \mR^M$, where $\mathop{\rm max}_X$ and $\mathop{\rm min}_{Y}$ stand for maximization and minimization with respect to $X$ and $Y$, respectively. $\rm T$ denotes the matrix transport. 

The objective function of eq. (\ref{unconstrained}) includes non-trivial interaction terms $\blambda^{\rm T}\bF \bx$, which is why computation cost is so considerable. Our key idea for reducing the cost is to approximate eq. (\ref{unconstrained}) by a bunch of single-body optimization problems that can be handled with a lower computational cost, which is similar to the spirit of mean field approximations of statistical mechanics \cite{OpperSaad2001}. 

For this, we introduce auxiliary functions $\phi_{i \to \mu}(x_i)$ and $\psi_{\mu \to i}(\lambda_\mu)$ to two types of optimization variables $x_i$ and $\lambda_\mu$, respectively. $\phi_{i \to \mu}(x_i)$ physically means the single-body objective function of $x_i$ for the ``$\mu$-cavity system'' that is defined by removing $\lambda_\mu$ from the original system, and similarly for $\psi_{\mu \to i}(\lambda_\mu)$ \cite{MezardParisiVirasoro1987,MezardMontanari2010}. The BP framework indicates that these functions can be determined by the following considerations: 
\begin{enumerate}
\item {\underline{$i$-cavity $\to$ $\mu$-cavity}}\\ 
Recover the original system by inserting $x_i$ to the $i$-cavity system. After that, choose an index $\mu \in {\cal M}(i)$, where ${\cal M}(i)$ denotes the set of indices of the Lagrange multipliers that are directly connected to $x_i$, and remove $\lambda_\mu$. This yields the $\mu$-cavity system. Optimizing the approximate objective function of the $\mu$-cavity system with respect to $\lambda_{\nu \in {\cal M}(i) \backslash \mu}$ offers the single body objective function of $x_i$ in the $\mu$-cavity system. Here, $A \backslash a$ denotes the set that is defined by removing an element $a$ from a set $A$. This provides
\begin{eqnarray}
&& \phi_{i \to \mu}(x_i)=|x_i| \cr
&& \phantom{\phi}+\sum_{\nu \in {\cal M}(i) \backslash\mu}
\mathop{\rm max}_{\lambda_\nu}
\left \{
F_{\nu i} \lambda_\nu x_i + \psi_{\nu \to i}(\lambda_\nu) \right \}. 
\label{h-step}
\end{eqnarray}

\item {\underline{$\mu$-cavity $\to$ $i$-cavity}}\\ 
Recover the original system by inserting $\lambda_\mu$ to the $\mu$-cavity system. After that, choose an index $i \in {\cal I}(\mu)$, where ${\cal I}(\mu)$ denotes the set of indices of the signal variables that are directly connected to $\lambda_\mu$, and remove $x_i$. This yields the $i$-cavity system. Optimizing the approximate objective function of the $i$-cavity system with respect to $x_{l \in {\cal I}(\mu) \backslash i}$ offers the single body objective function of $\mu$ in the $i$-cavity system. This provides
\begin{eqnarray}
&& \psi_{\mu \to i}(\lambda_\mu)=-y_\mu \lambda_\mu \cr
&& \phantom{\psi}+\sum_{l \in {\cal I}(\mu) \backslash i}
\mathop{\rm min}_{x_l}
\left \{
F_{\mu l} \lambda_\mu x_l + \phi_{l \to \mu}(x_l) \right \}. 
\label{v-step}
\end{eqnarray}
\end{enumerate}

Let us denote the single body objective functions for approximating eq. (\ref{unconstrained}) as
$\phi_i(x_i)$ and $\psi_\mu(\lambda_\mu)$. 
These can be constructed from the auxiliary functions as follows: 
\begin{enumerate}
\setcounter{enumi}{2}
\item \underline{Construction of $\phi_i(x_i)$ and $\psi_\mu(\lambda_\mu)$}\\
Recover the original system by inserting $\lambda_\mu$ to the $\mu$-cavity system and optimize the approximate objective function with respect to $\lambda_{\mu \in {\cal M}(i)}$. This yields
\begin{eqnarray}
\phi_i(x_i)\!=\!|x_i|\! +\!\!\!\sum_{\mu\in {\cal M}(i)}\!\!\!
\mathop{\rm max}_{\lambda_\mu}
\left \{
F_{\mu i} \lambda_\mu x_i \!+ \!\psi_{\mu \to i}(\lambda_\mu) \right \}. 
\label{phi_full}
\end{eqnarray}
Similarly, 
\begin{eqnarray}
\psi_\mu(\lambda_\mu)\!=\!-y_\mu \lambda_\mu \!\!+
\!\!\!\!\sum_{i\in {\cal I}(\mu)}\!\!\!\!
\mathop{\rm min}_{x_i}
\left \{
F_{\mu i} \lambda_\mu x_i \!+\! \phi_{i \to \mu}(x_i) \right \}. 
\label{psi_full}
\end{eqnarray}
\end{enumerate}
Under suitable conditions, iterating eqs. (\ref{h-step}) and (\ref{v-step}) for all connected pairs of $i$ and $\mu$ is expected to yield a convergent solution of $\phi_{i \to \mu}(x_i)$ and $\psi_{\mu \to i}(\lambda_\mu)$. Inserting the solution into (\ref{phi_full}) and optimizing $\phi_i(x_i)$ offer the recovered signal $\widehat{x}_i=\mathop{\rm argmin}_{x_i}\{\phi_i(x_i)\}$.

\subsection{Quadratic approximation}
As long as $j,k \sim O(1)$, the necessary computational cost for performing the above procedure grows linearly with $N$ since the optimization required at each step is concerned with only a small number of variables. However, the prefactor is considerably large since one has to update {\em functions} at each step, which may reduce practical applicability. This difficulty is shared with another BP-based algorithm developed by Baron et al. \cite{CSBP2006}. 

To reduce necessary computation cost, we limit $\phi_{i \to \mu}(x_i)$ and $\psi_{\mu \to i}(\lambda_\mu)$ to the form of (piecewise) quadratic functions as
\begin{eqnarray}
\phi_{i \to \mu}(x_i)&=&\frac{1}{2}A_{i \to \mu}x_i^2-B_{i \to \mu} x_i + |x_i|, 
\label{AB} \\
\psi_{\mu \to i}(\lambda_\mu)&=&-\frac{1}{2}C_{\mu \to i}\lambda_\mu^2+(D_{\mu \to i}-y_\mu)\lambda_\mu, 
\label{CD} 
\end{eqnarray}
and derive update rules for $A_{i \to \mu}$, $B_{i \to \mu}$, $C_{\mu \to i}$ and $D_{\mu \to i}$. For this, we first substitute eq. (\ref{CD}) into eq. (\ref{h-step}), which yields 
\begin{eqnarray}
&&A_{i \to \mu}=\sum_{\nu \in {\cal M}(i) \backslash \mu} \frac{F_{\nu i}^2}{C_{\nu \to i}}, 
\label{CDtoA} \\
&&B_{i \to \mu}=\sum_{\nu \in {\cal M}(i) \backslash \mu} \frac{F_{\nu i}}{C_{\nu \to i}}(y_{\nu}- D_{\nu \to i}), 
\label{CDtoB} 
\end{eqnarray}
exactly. Next, we insert eq. (\ref{AB}) into eq. (\ref{v-step}). In this, the expression 
\begin{eqnarray}
&&\mathop{\rm min}_{x_l} \left \{ F_{\mu l} \lambda_\mu x_l + \phi_{\mu \to l}(x_l) \right \}=\cr
&&\left \{
\begin{array}{ll}
-\frac{(B_{l \to \mu}-1-\lambda_\mu F_{\mu l})^2}{2 A_{l \to \mu}},\! & \! B_{l \to \mu}\!-\!\lambda_{\mu}\!F_{\mu l} 
\!>\!1, \cr
0, \!& \! |B_{l \to \mu}\!-\!\lambda_{\mu}\!F_{\mu l}|\!<\!1, \cr
-\frac{(B_{l \to \mu}+1-\lambda_\mu F_{\mu l})^2}{2 A_{l \to \mu}} 
\!&\! B_{l \to \mu}\!-\!\lambda_{\mu}\!F_{\mu l}\! <\!-1
\end{array}
\right .
\label{kubun}
\end{eqnarray}
should be paid attention to. For a fixed $\mu$, eq. (\ref{kubun}) represents a piecewise quadratic function that switches its functional form at two points $\lambda_\mu=(B_{l \to \mu}\pm 1)/F_{\mu l}$, which vary among $l \in {\cal I}(\mu)\backslash i$. This means that directly using eq. (\ref{kubun}) in assessing eq. (\ref{v-step}) yields a piecewise quadratic function that switches its functional form at $2(j-1)$ different points, which makes it impossible to obtain the quadratic form of eq. (\ref{CD}) any more. To practically resolve this problem, we approximate the right hand side of eq. (\ref{kubun}) by
\begin{eqnarray}
\left \{
\begin{array}{ll}
-\frac{(B_{l \to \mu}-1-\lambda_\mu F_{\mu l})^2}{2 A_{l \to \mu}}, &  B_{l \to \mu}
>1, \cr
0, &  |B_{l \to \mu}|<1, \cr
-\frac{(B_{l \to \mu}+1-\lambda_\mu F_{\mu l})^2}{2 A_{l \to \mu}}, 
& B_{l \to \mu} <-1
\end{array}
\right .
\label{kubun2}
\end{eqnarray}
which is justified if $|B_{l \to \mu}/F_{\mu l}| \gg 1$ holds. As eq. (\ref{kubun2}) represents not piecewise but totally quadratic functions of $\lambda_\mu$, using this approximation in assessing eq. (\ref{v-step}) yields the form of eq. (\ref{CD}), which leads to 
\begin{eqnarray}
&&C_{\mu \to i}=\sum_{l \in {\cal L}(\mu) \backslash i} F_{\mu l}^2
\frac{\partial f(B_{l \to \mu};A_{l \to \mu})}{\partial B_{l \to \mu}}, 
\label{ABtoC}\\
&&D_{\mu \to i}=
\sum_{l \in {\cal L}(\mu) \backslash i} F_{\mu l} f(B_{l \to \mu};A_{l \to \mu}). 
\label{ABtoD}
\end{eqnarray}
Here, $f(B;A)$ represents a soft-thresholding function 
\begin{eqnarray}
f(B;A)\equiv\left (B-\frac{B}{|B|} \right) \Theta(|B|-1)/A, 
\label{fBA}
\end{eqnarray}
in which $\Theta(u)=1$ for $u > 0$ and vanishes, otherwise. 

Under appropriate conditions, iterating eqs. (\ref{CDtoA}) and (\ref{CDtoB}) $\to $ eqs. (\ref{ABtoC}) and (\ref{ABtoD}) is expected to yield a convergent solution of $A_{i \to \mu}$, $B_{i \to \mu}$, $C_{\mu \to i}$ and $D_{\mu \to i}$. In addition, substituting eq. (\ref{CD}) into eq. (\ref{phi_full}) offers an expression
\begin{eqnarray}
\phi_i(x_i)=\frac{1}{2}A_i x_i^2-B_i x_i + |x_i|, 
\label{app1object}
\end{eqnarray}
where
\begin{eqnarray}
&&A_i=\sum_{\mu \in {\cal M}(i)} \frac{F_{\mu i}^2}{C_{\mu \to i}}, 
\label{fullA} \\
&&B_i=\sum_{\mu \in {\cal M}(i)} \frac{F_{\mu i}}{C_{\mu \to i}}
(y_\mu-D_{\mu \to i}). 
\label{fullB}
\end{eqnarray}
Equation (\ref{app1object}) provides the recovered signal as
\begin{eqnarray}
\widehat{x}_i=f(B_i;A_i). 
\label{recovered}
\end{eqnarray}
All this constitutes the main achievement of this article. Actual time required for running this algorithm is much less than that for the direct product from BP of eqs. (\ref{h-step}), (\ref{v-step}) and (\ref{phi_full}) because one only has to deal with four types of variables defined for each pair of signal variables and the Lagrange multipliers without handling their functions, although the growth rate of the computational cost with respect to $N$ is the same between the two algorithms. 

\subsection{Dense matrix limit}
To theoretically examine the property of the algorithm developed above, let us suppose an extreme situation where $j \to M$, $k \to N$ and matrix entries $F_{\mu i}$ independently follow an identical distribution of zero mean and variance $N^{-1}$. A consideration similar to the following has been provided in research on CDMA multiuser detection of wireless communication before \cite{Kabashima2003}. In the supposed situation, the law of large numbers suggests that $A_{i \to \mu} \simeq A_i \simeq A \equiv \alpha/C$ and $C_{\mu \to i} \simeq C \equiv N^{-1} \sum_{l=1}^N (\partial/\partial B_l)f(B_l;A)$ typically hold for $i=1,2,\ldots,N$ and $\mu=1,2,\ldots,M$, where we used an expression of the Taylor expansion
\begin{eqnarray}
&&  f(B_{l \to \mu};A)-f(B_{l};A) \simeq 
-\frac{\partial f(B_{l};A)}{\partial B_l} \frac{F_{\mu l}}{C}(y_\mu -D_{\mu \to l})\cr
&& \sim O(N^{-1/2}) \to 0 \ (N \to \infty). 
\label{Taylor}
\end{eqnarray}  
Further, inserting this and the expression of $y_\nu =\sum_{i=1}^N F_{\nu i}x_i^0$ into eq. (\ref{CDtoB}), we have 
\begin{eqnarray}
B_{i \to \mu}\simeq \frac{\alpha}{C}
x_i^0 +  \frac{1}{C}\sum_{\nu \ne \mu}
F_{\nu l} 
\sum_{l \ne i} F_{\nu l}\left (x_l^0-f(B_{l \to \nu};A) \right). 
\label{B_CLT}
\end{eqnarray}
Concerning this, the central limit theorem indicates that the second term of eq. (\ref{B_CLT}) converges to obey a Gaussian distribution of zero mean and variance $C^{-2} N^{-2} \sum_{\nu \ne \mu} \sum_{l \ne i} (x_l^0-f(B_{l \to \nu};A))^2 \simeq (\alpha/C^2) N^{-1} \sum_{l=1}^N (x_l^0-f(B_{l};A))^2$ and is independent among different $i$'s for given $\mu$ as $N$ tends to infinity since $F_{\mu i}$'s are independent of one another. 

These arguments mean that as $N,M=\alpha N \to \infty$, macroscopic variables with respect to the signal estimate of eq. (\ref{recovered}), $m=N^{-1} \sum_{i=1}^N x_i^0 \widehat{x}_i$ and $Q=N^{-1}\sum_{i=1}^N (\widehat{x}_i)^2$ are determined from the following equations: $m=\left \langle \int Dzx^0 f\left (B;A \right) \right \rangle_{x^0}$, $Q=\left \langle \int Dz \left(f\left (B;A \right) \right)^2\right \rangle_{x^0}$ and $C=\left \langle \int Dz \left (\partial f\left (B;A \right)/\partial B \right)\right \rangle_{x^0}$, where $B=(\alpha/C) x^0+(\sqrt{\alpha (Q-2m+Q_0)}/C)z$, $A=\alpha/C$ and $Dz=dz \exp (-z^2/2)/\sqrt{2 \pi}$ denotes the Gaussian measure. $\left \langle \cdots \right \rangle_{x^0}$ represents the average operation with respect to the original signal and $Q_0=\left \langle (x^0)^2 \right \rangle_{x^0}$. It may be noteworthy that these equations for determining the macroscopic variables are equivalent to those obtained by the replica method of statistical mechanics for the $l_1$-norm based signal recovery scheme \cite{KWT2009}. As the replica method reproduces 
the result identical to that obtained by mathematically rigorous analyses \cite{DonohoTanner2009,BayatiMontanari2010}, this suggests that the current algorithm can saturate a theoretical limit of the $l_1$-norm based scheme for the perfect recovery, which is often termed the Donoho-Tanner weak threshold \cite{DonohoTanner2009}, in the limit of dense matrices. 

The techniques based on the law of large numbers and the Taylor expansion, which are used above, are also useful for reducing the necessary computational cost in the dense matrix case. Inserting eq. (\ref{Taylor}) into $D_\mu \equiv \sum_{i=1}^N F_{\mu i} f(B_{i \to \mu};A)$ yields
\begin{eqnarray}
z_{\mu}=y_\mu-\sum_{i=1}^N F_{\mu i}\widehat{x}_i +
\frac{1}{C}\left (\frac{1}{N}\sum_{i=1}^N \frac{\partial f(B_i;A)}{\partial B_i} \right) z_\mu, 
\label{h-step_low}
\end{eqnarray}
where $z_\mu \equiv y_\mu -D_\mu$ and $\sum_{i=1}^N F_{\mu i}^2 \partial f(B_{i \to \mu};A)/\partial B_{i \to \mu}\simeq (1/N) \sum_{i=1}^N \partial f(B_{i};A)/\partial B_{i}$ was used in the right hand side on the basis of the law of large numbers. On the other hand, plugging $D_{\mu \to i}=D_\mu -F_{\mu i} f(B_{i \to \mu};A) \simeq D_\mu -F_{\mu i} \widehat{x}_i$ into eq. (\ref{fullB}) and using an effect of the law of large numbers $\sum_{\mu=1}^M F_{\mu i}^2 \simeq \alpha$ lead to 
\begin{eqnarray}
B_i=\frac{1}{C} \sum_{\mu=1}^M F_{\mu i} z_\mu + \frac{\alpha}{C} \widehat{x}_i. 
\label{v-step_low}
\end{eqnarray}
Iterating eqs. (\ref{h-step_low}) and (\ref{v-step_low}) in conjunction with $C=(1/N) \sum_{i=1}^N \partial f(B_{i};A)/\partial B_{i}$ and $A=\alpha/C$ offers the signal estimate $\widehat{x}_i =f(B_i;A)$ with an $O(MN)$ computational cost per update, which is considerably smaller than that of the original expression, $O(MN(M+N))$, when $N,M \gg 1$. 

In the context of the signal recovery of CS, a class of algorithms similar to eqs. (\ref{h-step_low}) and (\ref{v-step_low}), termed AMP, has already been proposed by Donoho et al. 
\cite{DonohoMalekiMontanari2009}. 
Performance of AMP generally depends on how a certain parameter for soft-thresholding, which corresponds to ``$A^{-1}$'' in the current algorithm, is controlled externally. A characteristic feature of the current algorithm is that this parameter is tuned adaptively for given $\bF$ and $\by$ in the course of updates. This property may be preferred in practical usage because one does not have to care about effects of sample fluctuations in such an adaptive control. 

\section{Experimental Validation}
To examine the abilities and limitations of the above scheme, we assessed the capability of the perfect signal recovery for cases (A) $(j,k)=(10,20)$ and (B) the dense matrix case by extensive numerical experiments. As they can be compared with accurate and mathematically rigorous assessments, we will first show the results for (B). 

\begin{figure}[t]
\setlength\unitlength{1mm}
\begin{picture}(100,55)(0,0)
\put(0,0){\includegraphics[width=9cm]{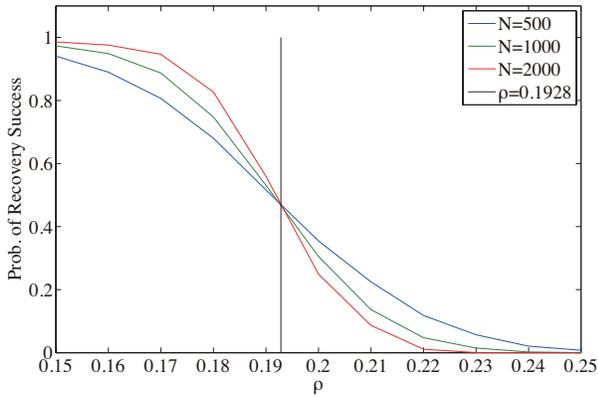}}
\end{picture}
\caption{Probability of perfectly recovering the original signal for dense matrices.}
\label{fig1}
\end{figure}

In the (B) experiment, the probability that the original signal is perfectly recovered 
up to 10000 iterations of eq. (\ref{h-step_low}) $\to$ $C=(1/N) \sum_{i=1}^N \partial f(B_{i};A)/\partial B_{i}$ $\to$ $A=\alpha/C$ $\to $ eq. (\ref{v-step_low}) $\to $ eq. (\ref{recovered}) was evaluated through 10000 trials for each pair of various signal density $\rho$ and signal length $N$. We focused on the case of $\alpha=M/N=1/2$. For each trial, we generated an $M(=N/2) \times N$ dense random matrix $\bF$, entries of which were sampled independently from an Gaussian distribution of zero mean and variance $1/N$. Each entry $x_i^0$ of the original signal $\bx^0$ was sampled from $P(x)=(1-\rho)\delta(x)+\rho \exp \left (-x^2/2 \right)$ independently. We judged that $\bx^0$ was perfectly recovered if $N^{-1} \sum_{i=1}^N (\widehat{x}_i-x_i^0)^2 < 10^{-8}$ is satisfied. The results for $N=$500, 1000, and 2000 are plotted in Fig. \ref{fig1}, which shows that three curves of the probability of the perfect recovery for the different system sizes intersect one another at a point very close to the Donoho-Tanner weak threshold for $\alpha=1/2$, $\rho_{\rm c}(1/2)=0.1928\ldots $. This suggests that the developed algorithm can saturate the theoretical limit of the perfect recovery in the dense matrix case with a computational cost of $O(N^2)$ per update, which is lower than that required for the generic interior point method. 

\begin{figure}[t]
\setlength\unitlength{1mm}
\begin{picture}(100,110)(0,0)
\put(0,55){\includegraphics[width=9cm]{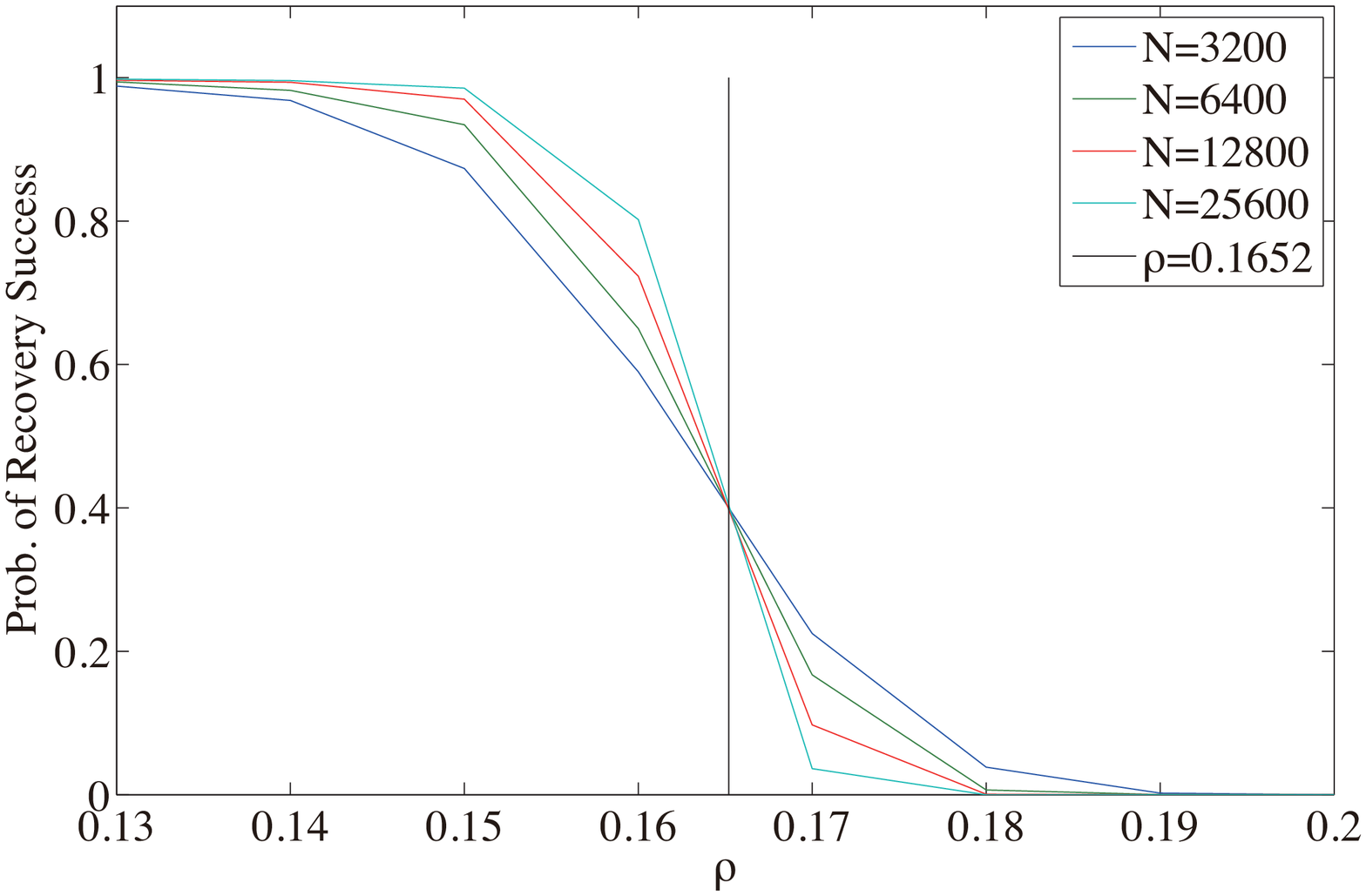}}
\put(5,105){(a)}
\put(0,0){\includegraphics[width=9cm]{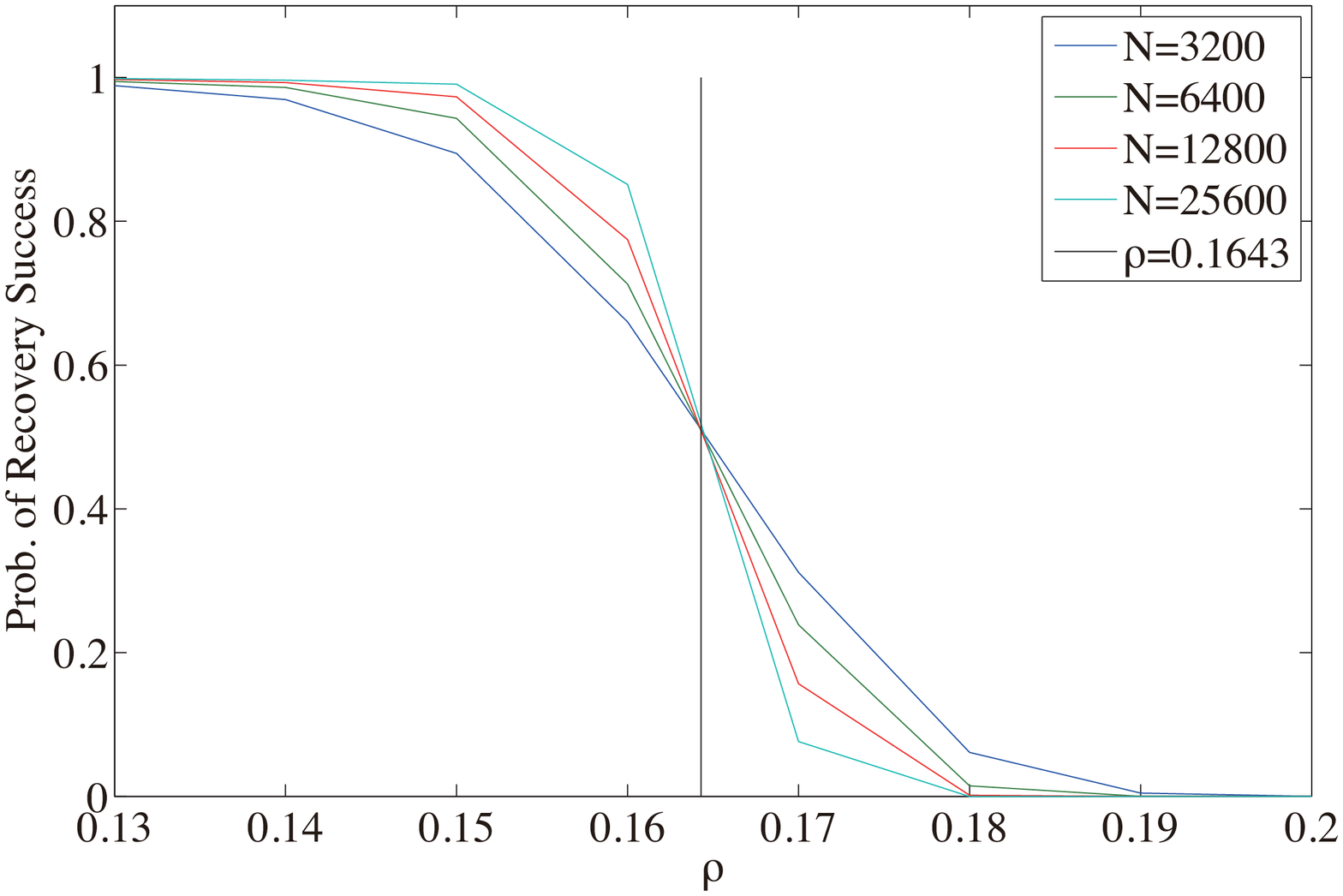}}
\put(5,48){(b)}
\end{picture}
\caption{(a): Probability of perfectly recovering the original signal evaluated for randomly constructed sparse matrices of $(j,k)=(10,20)$. (b): That assessed by the ``density evolution'' scheme (see the main text for details).}
\label{fig2}
\end{figure}

For the case of (A), we also assessed the probability of the perfect recovery by using iterations of eqs. (\ref{CDtoA}) and (\ref{CDtoB}) $\to $ eqs. (\ref{ABtoC}) and (\ref{ABtoD}) for sparse matrices which were randomly constructed under the constraint of $(j,k)=(10,20)$. Values of non-zero entries were independently sampled from the Gaussian of zero mean and unit variance. As convergence is rather faster than in the dense matrix case, we set the maximum number of iterations to 1000. The ways of generating $\bx^0$, the condition for judging the perfect recovery, and the number of trials for each parameter setting were the same as the above. Figure \ref{fig2} (a) plots the results for $N=$3200, 6400, 12800, and 25600 and shows that four curves for the different system sizes non-trivially intersect one another at $\rho \sim 0.1652$. The accurate estimate for the theoretical limit of the perfect recovery has not been clarified for the sparse matrices. Alternatively, we performed another experiment by shuffling connectivities among variables at each iteration, which corresponds to the density evolution analysis of BP \cite{RichardsonUrbanke}. The results are presented in Fig. \ref{fig2} (b). A non-trivial cross of four curves is also observed at $\rho \sim 0.1643$, where the difference in the third digit from that in Fig. \ref{fig2} (a) may be attributed to effects of finiteness of the system sizes and the maximum number of iterations. These suggest that for the sparse matrices the algorithm developed here can typically recover the original signal perfectly with an $O(Nk+Mj)$ cost of computations per update if the density of non-zero entries in the signal is below a certain finite critical value $\rho_{\rm th}(\alpha)$ as $N,M \to \infty$ keeping $\alpha =M/N \sim O(1)$.

\section{Summary}
In summary, we have developed an approximate signal recovery algorithm with low computational cost for sparse matrix based compressed sensing. The developed algorithm saturates the Donoho-Tanner weak threshold for the perfect recovery when the measurement matrix becomes dense. For sparse matrices, the algorithm is still capable of typically recovering the original signal perfectly if the density of the non-zero entries in the original signal is lower than a certain finite critical value as the signal length and the number of measurements tends to infinity while keeping the ratio between them finite. 

Accurately identifying the theoretical limit of the perfect signal recovery for the sparse matrix based compress sensing
and designing sparse matrices \cite{Dimakis2010} for improving the performance of the 
current algorithm are challenging problems for future studies. 

\section*{Acknowledgments}
The authors thank KAKENHI Nos. 22300003, 22300098 (YK), 22560370 (TW) and The Mitsubishi Foundation (YK) for their financial support. The authors also acknowledge the JSPS GCOE ``CompView'' for letting them use the TSUBAME Computing Services of Tokyo Tech. 
YK also appreciates useful discussions with M. M\'{e}zard at a preliminary stage of this work. 


%

\end{document}